\documentclass[%
 reprint,
superscriptaddress,
 amsmath,amssymb,
 aps,
prl]{revtex4-2}

\usepackage{amssymb}
\usepackage{graphicx}% Include figure files
\usepackage{dcolumn}% Align table columns on decimal point
\usepackage{bm}% bold math
\usepackage{hyperref}
\usepackage{xcolor}
\usepackage{tabularx}
\usepackage{booktabs}
\usepackage{multirow}
\usepackage{svg}
\usepackage{bibunits}
\defaultbibliography{Bibliography}
\defaultbibliographystyle{apsrev4-2}
\makeatletter
\@namedef{b@apsrev42Control}{}
\makeatother
\newcommand{\tb}{\textcolor{blue}}
\newcommand{\cH}{\mathcal{H}}
\newcommand{\bq}{\mathbf{q}}
\newcommand{\Mext}{\mathrm{M}_{\mathrm{ext}}}
\newcommand{\TCW}{\theta_{\text{CW}}}
\newcommand{\SF}{\tilde{S}}
\newcommand{\bk}{\mathbf{k}}
\newcommand{\ns}{\hat{\boldsymbol{\sigma}}_n}
\newcommand{\DDCS}{\frac{\mathrm{d}^2 \sigma}{\mathrm{d} \epsilon \mathrm{d} \Omega}}
\newcommand{\Sq}{\hat{\mathbf{S}}_{\mathbf{q}}}

\newcommand{\Sperp}{\hat{\mathbf{S}}_{\mathbf{q}}^\perp}

\begin{document}
\begin{bibunit}
\nocite{apsrev42Control}

\preprint{APS/123-QED}

\title{Spin correlations, low-energy scales, and anisotropy scaling \\ in 2D frustrated magnets}% Force line breaks with \\

% Identifying low-energy spectral features in frustrated magnets via adiabatic continuity 

\date{\today}% It is always \today, today,

\begin{abstract}
Neutron scattering is central to identifying quantum states of magnetic materials. In the search for quantum spin liquids, broad spectral features of inelastic spectra have been cited as evidence for spinon excitations, but can also arise from magnon excitations in the presence of quenched disorder and strong magnon interactions.
We develop a new approach to this problem, based on the adiabatic continuity in the 
$XXZ$ Heisenberg model on geometrically frustrating (GF) lattices as a function of the model's anisotropy.
Using this approach, we identify universal features and energies of finite-temperature spin correlators.
Focusing on the kagome, triangular $J_1 - J_2$, and square-kagome lattices, we show that the low-energy spin spectral functions exhibit robust peaks. These peaks are shown to be closely tied to the ``hidden energy scale'' $T^*$, the characteristic scale of a low-temperature peak in the heat capacity, at which many GF magnets also display spin-glass freezing. We show that
the spectral features at low energies $\omega\lesssim J$
arise from single-magnon scattering and identify the magnetizations of the respective excitations. For the kagome lattice, we also explore the evolution of the spectral features with temperature. Our results provide 
a sharp spectroscopic criterion for interpreting neutron scattering in a broad class of GF quantum magnets.
\end{abstract}

\author{Phillip Popp}
\affiliation{Physics Department, University of California, Santa Cruz, California 95064, USA}

\author{Stephan Rosenkranz}
\affiliation{Materials Science Division, Argonne National Laboratory, Lemont, Illinois 60439, USA}

\author{Arthur P. Ramirez}
\affiliation{Physics Department, University of California, Santa Cruz, California 95064, USA}

\author{Sergey Syzranov}
\affiliation{Physics Department, University of California, Santa Cruz, California 95064, USA}

\maketitle

Neutron scattering (NS) is an extremely powerful tool for investigating magnetic correlations in materials and can reveal the existence and nature of magnetic order as well as the nature of elementary excitations. Of relevance here is that inelastic scattering features have been used to argue for the existence of theoretically predicted quantum-spin-liquid (QSL) states.

Conventional magnetic states, with well-defined magnon excitations, display sharp peaks in the spectral weight vs. energy/momentum, while broad continua may hint at 
the highly sought-after QSLs~\cite{Balents:SL,SavaryBalents:QSL}. Such ``particle continua'' are often interpreted~\cite{ColdeaTennantTylczynski:Spinons,Helton:SRO,KohnoStarykhBalents:Spinons,Han:Herbertsmithite1,Han:Herbertsmithite2,Shen:Spinons1,Gao:Pyrochlore,Smith:Pyrochlore,Zeng:Spinons,Park:BLCTO,Pore:Pyrochlore,Thennakoon:KHAF,Breidenbach:Zn-Barlowite} as evidence of fractionalized excitations, spinons, that may be produced in pairs in neutron-scattering processes.

Broad spectral-weight features may also come from a variety of mundane sources, such as impurities,
quasispins~\cite{SchifferDaruka:Quasispins,LaForge:Quasispins,RamirezSyzranov:SRO,Sedik:Quasispins}, and magnons, when the magnon self-energy is broadened by quenched disorder and/or interactions~\cite{Matan:Jarosite1,Christensen:Entanglement,PerkinsBrenig:THAF,Mourigal:THAF,Oh:THAF,ChenQu:THAF,Chaebin:THAF}. 
Identifying the origins of neutron-scattering signals is, therefore, critical for the determination of the nature of the magnetic states and investigating the properties of magnetic excitations.

In this Letter, we identify universal spectral features of spin fluctuations and relations between
energy scales in geometrically frustrated materials (GFMs), the largest class of 
candidate materials for QSLs. 
We develop a new approach for investigating such 
features, which is based on
the adiabatic continuity of the properties of excitations in the $XXZ$ Heisenberg model with the Hamiltonian
\begin{equation}
\hat{\cH}_{XXZ} = \sum_{(ij)} J_{ij} \biggl{(} \hat{S}_i^{z} \hat{S}_j^z + \alpha \biggl{\{} \hat{S}_i^x \hat{S}_j^x + \hat{S}_i^y\hat{S}_j^y\biggr{\}} \biggr{)}
\label{eq:HXXZ}
\end{equation}
% \begin{equation}
% \hat{\cH}_{XXZ} = J \sum_{(ij)} \biggl{(} \hat{S}_i^{z} \hat{S}_j^z + \frac{\alpha}{2} \biggl{\{} \hat{S}_i^+ \hat{S}_j^- + \hat{S}_i^- \hat{S}_j^+ \biggr{\}} \biggr{)}
% \label{eq:HXXZ}
% \end{equation}
when tuning the ratio $\alpha$ of the transverse to longitudinal coupling.

%%%%%%%%%%%%%%%%%%%%%%%%%%%%%%%%%%%%%%%%%%%%%%%%%%%%%%%%%%%%%%%
%%%%%%%%%%%%%%%%%%%%%%%%%%%%%%%%%%%%%%%%%%%%%%%%%%%%%%%%%%%%%%%
\begin{figure*}[t]
    \centering
    \includegraphics[width = 0.94\textwidth]{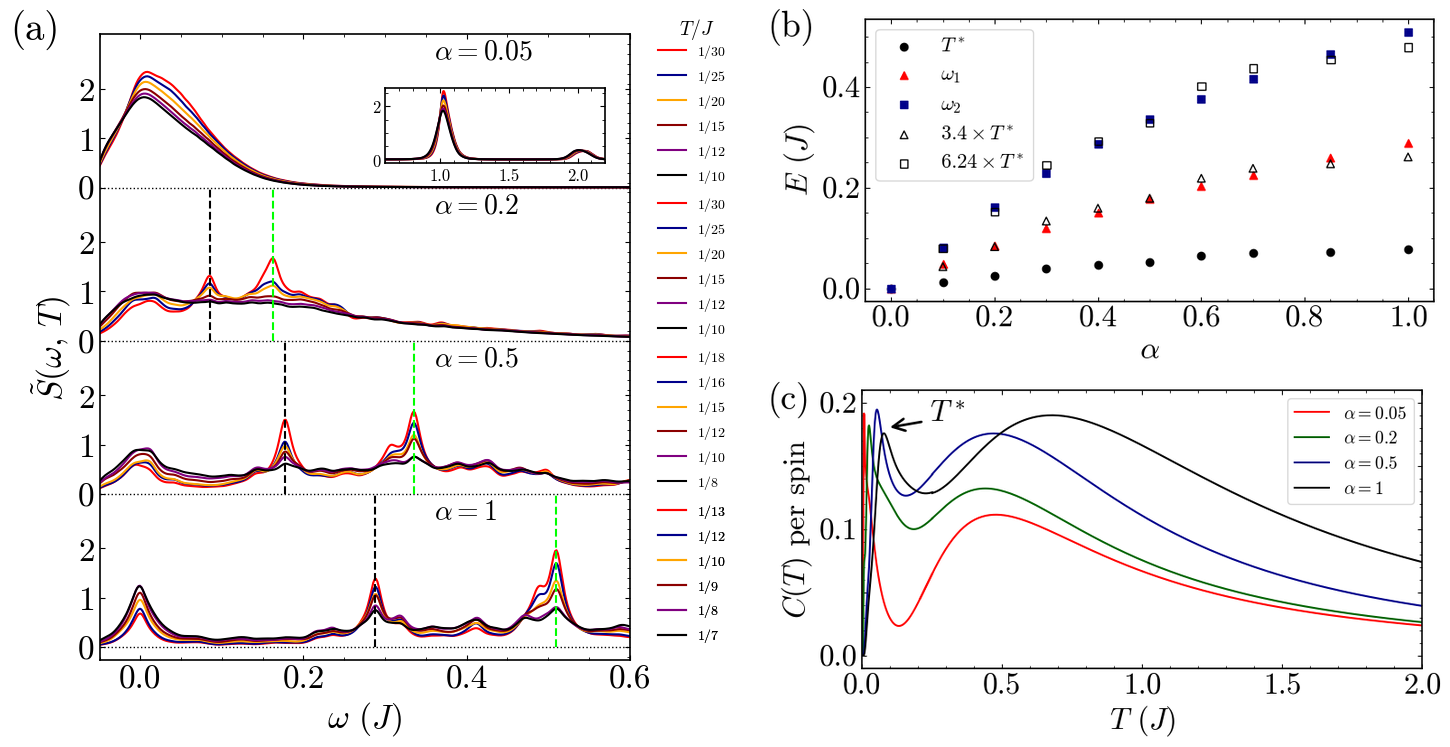}
    \caption{{\bf (a)} 
    The transverse dynamical spin structure factor $\SF (\omega, \, T)$ of $N = 24$ kagome clusters
    having various ratios $\alpha$ of the transverse to longitudinal
    exchange couplings and temperatures.
  Inset shows the structure factor at higher frequency for $\alpha = 0.05$. Dashed black (lime) lines indicate the position of $\omega_{1 (2)} (\alpha)$. {\bf (b)} Dependence of the three energy scales $T^*$, $\omega_1$, and $\omega_2$ on $\alpha$ (cf. panel {\bf c}). Open symbols denote fits to $\omega_{1(2)} (\alpha) = A_{1(2)} T^* (\alpha)$. {\bf (c)} Heat capacity per spin of $N = 24$ kagome clusters at the indicated values of $\alpha$, computed by the finite-temperature Lanczos method. The position of the low-temperature peak is denoted $T^* (\alpha)$.}
    \label{fig:S_xx_Summary}
\end{figure*}
%%%%%%%%%%%%%%%%%%%%%%%%%%%%%%%%%%%%%%%%%%%%%%%%%%%%%%%%%%%%%%%
%%%%%%%%%%%%%%%%%%%%%%%%%%%%%%%%%%%%%%%%%%%%%%%%%%%%%%%%%%%%%%%

Such a model continuously connects the isotropic Heisenberg model at $\alpha=1$, realistic for a great variety of materials,
to the significantly more analytically tractable spin ice at $\alpha\ll1$, with more easily-identifiable excitations.
Additionally, the dependencies of energy scales of spin correlations on $\alpha$ reveal hidden connections between them that cannot be established in neutron scattering experiments and numerical studies of fixed-$\alpha$ Heisenberg models. 

\tb{Summary of results.} 
%The main focus of this Letter is the kagome lattice, which belongs to some of the most intensively investigated QSL candidates, exemplified by Herbertsmithite~\cite{Helton:SRO,deVries:Herbertsmithite,Helton:Susceptibility,Han:Herbertsmithite1,Han:Herbertsmithite2,Jiang:KHAM}, and is also a subject of abundant numerical studies~\cite{Elser:KHAF,ZengElser:KHAF,ElstnerYoung:KHAF,NakamuraMiyashita:KHAF,TomczakRichter:KHAF,WaldtmannEverts:KHAF,SindzingreMisguich:KHAF,MisguichBernu:KHAF,MisguichSindzingre:KHAF,Laeuchli:DSSF, IsodaNakano:XXZ,YanSimengHuse:DMRG,DepenbrockMcCullochSchollwock:DMRG,Munehisa:KHAF,Shimokawa:SSF,Sherman:SF, SchnackSchulenberg:KHAF,Zhu:DSSF,HalimehSingh:KHAF,Zhang:DSSF,Prelovsek:DSSF,Ulaga:EasyAxis,Jiang:KHAM}.
The bulk of this Letter is devoted to the kagome lattice, which provides a self-contained illustration of our arguments. To demonstrate the universality of our main result, we also present computations for the $J_1 - J_2$ triangular and square-kagome lattices. We argue that the obtained qualitative trends should be expected generically for GFMs.

For each of these lattices, we show that the finite-temperature dynamical spin structure factor (DSSF) exhibits peaks at frequencies $\omega \ll J$ corresponding to transitions between low-lying magnetization sectors. We further establish that
these frequencies are tied to the so-called ``hidden energy scale''
$T^*$~\cite{SyzranovRamirez:EP,RamirezSyzranov:SRO,PoppRamirezSyzranov:HES}.
As established in earlier work, the heat capacity of strongly geometrically frustrated magnets exhibits two peaks~\cite{Elser:KHAF,ZengElser:KHAF,ElstnerYoung:KHAF,NakamuraMiyashita:KHAF,TomczakRichter:KHAF,SindzingreMisguich:KHAF,MisguichBernu:KHAF,MisguichSindzingre:KHAF,IsodaNakano:XXZ,SugiuraShimizu:KHAF,Munehisa:KHAF,Shimokawa:SSF,SchnackSchulenberg:KHAF,PrelovsekKokalj:THAF,ChenQu:THAF,MoritaTohyama:KagomeTriangular,SekiYunoki:RingExchange,Ulaga:EasyAxis,Hutak:Hyperkagome,PoppRamirezSyzranov:HES}, with the lower and higher peaks located, respectively,
at the scale $T^*$ and the Curie-Weiss temperature $\theta_\text{CW}$. Due to the adiabatic connection of the Hamiltonian~\eqref{eq:HXXZ} at $\alpha > 0$ to the Ising model at $\alpha= 0$, the entropy held by the lower peak matches the Ising ground-state entropy of the corresponding lattice~\cite{Nakatsuji:THAF,RamirezSyzranov:SRO,PoppRamirezSyzranov:HES,KrokhmalskiiHutakDerzhko:SpecificHeat,ZhuRamirezSyzranov:Entropy}. As revealed in Ref.~\cite{SyzranovRamirez:EP},
many frustrated magnets exhibit spin-glass freezing
at temperatures $T\lesssim T^*$, as well as increased intensity of neutron scattering.

The peak frequencies on each lattice and their relations to $T^*$ are summarized in Table~\ref{tab:frequency_summary}. These relations persist for arbitrary $\alpha$, so long as the corresponding system avoids magnetic long-range order.

\begin{table}[h]
    \centering
    \begin{ruledtabular}
    \begin{tabular}{ccc}
         Lattice & Low-frequency inelastic peak(s) \\
         \hline
         Kagome & 
         $\omega_1 = 3.4 T^*, \,\, \omega_2 = 6.24 T^*$ & \refstepcounter{equation}(\theequation)\label{eq:kagome_peaks} \\
         Triangular $J_1 - J_2$ & $\omega_{\text{t}} = 9.3 T^*$ & \refstepcounter{equation}(\theequation)\label{eq:triangular_peak} \\
         Square-kagome & $\omega_{\text{sk}} = 2.56 T^*$ & \refstepcounter{equation}(\theequation)\label{eq:sk_peak} \\
    \end{tabular}
    \end{ruledtabular}
    \caption{Frequencies of low-energy spectral peaks and their relations to $T^*$ for the models considered in this Letter.}
    \label{tab:frequency_summary}
\end{table}

% On the kagome lattice, at temperatures $T$ below $T^*$, the area under both peaks grows rapidly, indicating growing thermal occupation of states with the respective energies. At $T = 0$, the peak at $\omega_1$ disappears~\cite{SM}, as it arises from transitions between excited states. At sufficiently large $T$, both peaks disappear.

%%%%%%%%%%%%%%%%%%%%%%%%%%%%%%%%%%%%%%%%%%%%%%%%%%%%%%%%%%%%%%%%
%%%%%%%%%%%%%%%%%%%%%%%%%%%%%%%%%%%%%%%%%%%%%%%%%%%%%%%%%%%%%%%%

\tb{Simulation details.}
We use the low-temperature Lanczos method~\cite{Aichhorn:LTLM,Prelovsek:FTLM,Kokalj:Thesis} (LTLM)
to compute the DSSF:
\begin{equation}
S_{\bq}^{\gamma \delta} (\omega, \, T) = \frac{1}{2 \pi} \int_{-\infty}^\infty \mathrm{e}^{i \omega t} \bigl{\langle} \hat{S}_\bq^{\gamma \dagger} (t) \hat{S}_\bq^\delta (0) \bigr{\rangle}_T \, \text{d} t,
\label{eq:DSSF}
\end{equation}
for spins-$1/2$,
where $\hat{S}_\bq^{\gamma} (t)$ is the lattice Fourier transform of the $\gamma$-component of the local Heisenberg spin operator; and $\langle \dots \rangle_T \equiv \mathrm{Tr\left(e^{-\hat{\cH}/T} \ldots \right)}\left/ \mathrm{Tr\left(e^{-\hat{\cH}/T}\right)}\right.$denotes thermal averaging at temperature $T$. In spin-polarized neutron-scattering experiments on single crystals, the measured differential cross-section can be resolved into spin-flip and non-spin-flip channels that are proportional to the transverse and longitudinal components of~\eqref{eq:DSSF}, respectively.

We compute the transverse component of the correlator~\eqref{eq:DSSF}, $S_{\bq}^{xx}(\omega, \, T)=S_{\bq}^{yy}(\omega, \, T)$. At $\alpha = 1$, these correlators
match $S_{\bq}^{zz}(\omega, \, T)$. By contrast, for $\alpha \ll 1$, $\hat{S}_{\bq}^{z}$ does not generate excitations of interest. Thus, we focus on the transverse spin components.

\tb{Kagome lattice.} Our main focus here, the kagome lattice belongs to some of the most intensively investigated QSL candidates, exemplified by Herbertsmithite~\cite{Helton:SRO,deVries:Herbertsmithite,Helton:Susceptibility,Han:Herbertsmithite1,Han:Herbertsmithite2,Jiang:KHAM}, and is also a subject of abundant numerical studies~\cite{Elser:KHAF,ZengElser:KHAF,ElstnerYoung:KHAF,NakamuraMiyashita:KHAF,TomczakRichter:KHAF,WaldtmannEverts:KHAF,SindzingreMisguich:KHAF,MisguichBernu:KHAF,MisguichSindzingre:KHAF,Laeuchli:DSSF, IsodaNakano:XXZ,YanSimengHuse:DMRG,DepenbrockMcCullochSchollwock:DMRG,Munehisa:KHAF,Shimokawa:SSF,Sherman:SF, SchnackSchulenberg:KHAF,Zhu:DSSF,HalimehSingh:KHAF,Zhang:DSSF,Prelovsek:DSSF,Ulaga:EasyAxis,Jiang:KHAM}.
The three-site basis of the kagome structure means that NS intensity is not periodic over the elementary Brillouin zone (BZ) set by the Bravais lattice vectors, but rather an extended BZ with twice the linear dimensions. Because experiments~\cite{Helton:SRO,Han:Herbertsmithite1,Han:Herbertsmithite2,Thennakoon:KHAF,Breidenbach:Zn-Barlowite,Jiang:KHAM}, prior simulations~\cite{Laeuchli:DSSF,Zhu:DSSF,Zhang:DSSF,Shimokawa:SSF,Sherman:SF,Prelovsek:DSSF} and simulations reported in this work find broad low-energy spectral weight along the extended-BZ edges, maximized at the corresponding M-points, $\Mext$, we focus on computing spin correlators at $S_{\bq = \Mext}^{xx} (\omega, \, T) \equiv \SF (\omega, \, T)$. However, we show in Supplemental Material~\cite{SM} that the peak frequencies~\eqref{eq:kagome_peaks} are nearly dispersionless.
Though we carry out most simulations on a $24$-site cluster, the essential features of the DSSF are reproduced on $30$-site clusters (see Supplemental Material~\cite{SM}).

The key results for the kagome DSSF are presented in Fig.~\ref{fig:S_xx_Summary}a.
In what immediately follows, we describe the characteristic features in various energy ranges.

\tb{High-energy excitations ($E\gtrsim J$).}
In the Ising limit, $\alpha = 0$, the energies of all such excitations are multiples of $J$. Due to the adiabatic continuity of the excitations, 
the correlators, therefore, exhibit peaks at energies close to multiples of $J$
at $0 < \alpha \ll 1$,
as shown in the inset of Fig.~\ref{fig:S_xx_Summary}a for $\alpha = 0.05$. The correlator at small $\alpha$ has peaks close to $\omega=J$ and $\omega=2J$, corresponding to higher-energy Ising-like spin-flips.

% -----------------------------------------------------------------
%  On the other hand, in such an eigenbasis, $\hat{S}_\bq^x$ only has nonzero matrix elements between states that satisfy $|\Delta M_z| = 1$. Thus, at temperatures $T \lesssim T^*$, $\SF (\omega, \, T)$ can only be nonzero at frequencies close to the energies of single spin-flips from Ising ground states. As shown in Fig.~\ref{fig:Ising_SF}, the three possible energies of such spin-flips on the kagome lattice are $0, \, J$, and $2J$, precisely the energy scales seen in Fig.~\ref{fig:S_xx_Summary}a. 
%  ----------------------------------------------------------------

At all values of $\alpha$, the excitations with the energies $E\sim J$ are adiabatically connected to their Ising counterparts and give rise to the respective 
peak of the heat capacity~\cite{PoppRamirezSyzranov:HES,RamirezSyzranov:SRO}
at $\TCW$, the characteristic energy of spin-flip-type excitations.

%----------------------------------------------------------

% \tb{Small-$\alpha$ results.} For $0 < \alpha \ll 1$, we find that in a broad range of $T \lesssim 0.15 J$, $S_{\Mext}^{xx} (\omega, \, T)$ displays three distinct peaks at $\omega \approx 0, \, J$, and $2J$, respectively. 

% This is illustrated in Fig.~\ref{fig:S_xx_Summary}a for $\alpha = 0.05$. Notably, while the three-peak structure is robust across all simulated temperatures, the heights of the peaks at $\sim 0, \, J$ increase as $T$ decreases, whereas the height of the peak at $\sim 2J$ decreases slightly with $T$. We further discuss this temperature-dependence later, alongside the results for larger $\alpha$.

% ----------------------------------------------------------

\tb{Low-energy excitations ($E\ll J$).}
The excitations with low energies $E\ll J$ are superpositions of the Ising ground states and give rise to the lower heat capacity peak at the hidden energy scale~\cite{PoppRamirezSyzranov:HES,RamirezSyzranov:SRO}. Neutron scattering gives further insight into the structure of such states. 

 % Indeed, on one hand, the $z$-component $M_z$ of the total magnetization is a good quantum number of the Hamiltonian~\eqref{eq:HXXZ} for all values of $\alpha$. Thus, for $0 < \alpha \ll 1$, there exists an energy eigenbasis in which the lowest-lying excitations of~\eqref{eq:HXXZ}---which have energies $\sim T^* \sim E_{\mathrm{exc.}}$---are superpositions of Ising ground states that are also eigenstates of $\hat{M}_z$.

At small $\alpha\lesssim 0.1$ the features of the low-energy excitations merge into a single peak in $\SF (\omega, \, T)$ at $\omega=0$.
For $0.1 \lesssim \alpha \leq 1$, three peaks of the spectral weight of low-energy excitations can be resolved, as shown in Fig.~\ref{fig:S_xx_Summary}a.
The peak centered at $\omega=0$ corresponds to elastic neutron-spin scattering, while the other two peaks, at frequencies 
$\omega_1$ and $\omega_2$ [cf. Eq.~\eqref{eq:kagome_peaks}] come from inelastic scattering. Outside of the main peaks, there are weaker continua of nonzero intensity, similar to those reported in Refs.~\cite{Laeuchli:DSSF,Zhu:DSSF,Zhang:DSSF,Shimokawa:SSF,Sherman:SF,Prelovsek:DSSF}. 

%\begin{figure}[t]
%    \centering
%    \includegraphics[width=0.34\textwidth]{figure_2.png}
%    \caption{Three possible energies of flipping a single spin in a ground state of %the kagome-lattice Ising model. Filled (open) circles represent spin-up (down) %states. 
    %%The corresponding spin values are taken to be $\pm \frac{1}{2}$.
    %\label{fig:Ising_SF}
    %}
%\end{figure}

\begin{figure}[t]
    \centering
    \includegraphics[width=0.45\textwidth]{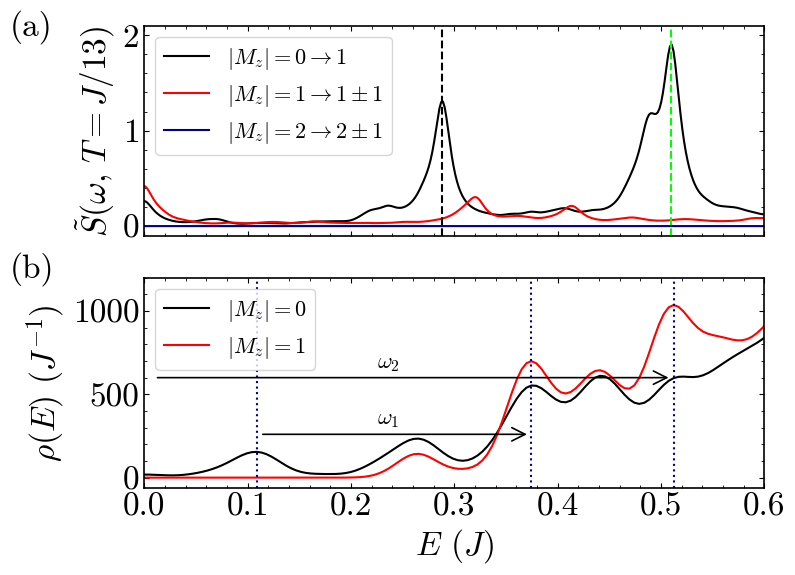}
    \caption{{\bf (a)} Contributions to $\SF (\omega, \, T = J / 13)$ of the $\alpha = 1$, $N = 24$ kagome cluster from the three lowest-magnetization transition channels. Dashed black (green) line indicates the position of $\omega_{1(2)}$. {\bf (b)} Low-energy $|M_z| = 0$ and $|M_z| = 1$ densities of states for an $\alpha = 1$ cluster.} %Dotted blue lines indicate the positions of peaks that are relevant for the DSSF; arrows indicate the associated transition frequencies.}
    \label{fig:Magnetization_Resolved}
\end{figure}

\tb{Relations between energy scales via $\alpha$-scaling.}
Our simulations of the correlator $\SF (\omega, \, T)$ [Fig.~\ref{fig:S_xx_Summary}a] show that at small $0<\alpha\ll1$, the energy scales $\omega_1$ and $\omega_2$ scale linearly with $\alpha$ [Fig.~\ref{fig:S_xx_Summary}b], indicating that 
these energy scales come from transitions involving excitations 
that emerge in the first-order (degenerate) perturbation theory in the transverse coupling between spins.

The heat capacity $C(T)$ shows two distinct peaks [Fig.~\ref{fig:S_xx_Summary}c]. The temperature $T^*$ of the lower peak also scales linearly for small $\alpha$ [see Fig.~\ref{fig:S_xx_Summary}b].

The relations between the energies $\omega_1$ and $\omega_2$
and the scale $T^*$ satisfy Eq.~\eqref{eq:kagome_peaks}.
Our results [Fig.~\ref{fig:S_xx_Summary}b] show, however, that
those relations are not specific to small $\alpha$ but persist up to
$\alpha=1$.
These universal relations let one identify spectral-weight peaks in neutron-scattering experiments in materials with arbitrary interaction anisotropy.

% For small $\alpha$, we find that the transverse DSSF at $\Mext$, the M-point of the extended Brillouin zone (BZ), displays peaks at the energies of certain spin flips in the corresponding Ising system. For larger values of $\alpha$, including the isotropic Heisenberg system, we find that the inelastic response exhibits two low-energy scales in the form of distinct peaks at frequencies $\omega_1^* (\alpha)$ and $\omega_2^* (\alpha)$. We show that the $\alpha$-dependence of both frequencies is consistent with Eq.~\eqref{eq:HES_Estimate}, and that both frequencies arise from the same magnetization transition channel. Further, we show that the temperature dependence of the integrated low-energy intensity mirrors the experimental trends.
% \tr{One has to sell results much stronger.}

%%%%%%%%%%%%%%%%%%%%%%%%%%%%%%%%%%%%%%%%%%%%%%%%%%%%%%%%%%%%%%%%

\tb{Magnetizations of low-energy excitations.} 
The magnetization $M_z$ of excitations is a good quantum number
in the $XXZ$ model under consideration.
To gain further insight into the nature of the origin of the $\SF(\omega, \, T)$ peaks, we analyze the magnetizations of the excitations contributing to those peaks.

Only the transitions changing the magnetization by $1$, $M_z\rightarrow M_z\pm 1$, contribute
to the simulated transverse correlator $\SF(\omega, \, T)=S^{xx}_\bq(\omega, \, T)$. 
We show the leading contributions in Fig.~\ref{fig:Magnetization_Resolved}a for $\alpha = 1$ and $T = J / 13$. 
We find that, to a good accuracy, 
the  peaks at $\omega_1$ and $\omega_2$ arise entirely from the
$M_z = 0 \rightarrow \pm 1$ channel, while the elastic peak (at $\omega=0$) comes from the $M_z = 0 \rightarrow\pm1$ and $|M_z|= 1 \rightarrow 1 \pm 1$ contributions.
Contributions from higher-$|M_z|$ channels are negligible.
%at the considered temperature.

\tb{Density of states and origins of peaks.}
We compute the magnetization-resolved density of states $\rho (E)$, with the results shown in Fig.~\ref{fig:Magnetization_Resolved}b. Both the $|M_z| = 0$ and $|M_z| = 1$ states exhibit several bands (with bands or band edges corresponding to peaks in $\rho (E)$). 
The higher-energy $|M_z| = 1$ peak lies at $\omega_2$, thus the associated transverse-DSSF peak comes from the transitions between the $M_z = 0$ ground state to the higher $|M_z|=1$ band, as shown in Fig.~\ref{fig:Magnetization_Resolved}b and verified by computing the appropriate matrix elements of $\hat{S}_\bq^x$.
The energy scale $\omega_1$ matches the energy difference between the lower $|M_z|=1$ peak and the first $M_z=0$ peak,
demonstrating that the corresponding DSSF peak comes from transitions between those bands. While a peak originating from transitions from the ground state to the lower $|M_z| = 1$ peak is not observed, the corresponding spectral signature may be suppressed by sublattice interference.

%Taken together, the $|M_z|$-resolved DSSF and density of states connect our results to theoretical QSL models and experiments. %The leading theoretical candidates are a gapless $U(1)$ Dirac QSL~\cite{Ran:DSL,Hermele:DSL} and a gapped $\mathbb{Z}_2$ %QSL~\cite{YanSimengHuse:DMRG,DepenbrockMcCullochSchollwock:DMRG,Sachdev:QMC,Poilblanc:RVB,Schuch:RVB,LuRanLee:Z2}, with %competing numerical support for both. 
%\tr{I would not like to over-advertise a $Z_2$ QSL or to say that the $M_z$ excitations are visons or that we see a two-spinon continuum.
%Maybe those are sharp excitations that have some dispersion.}
%However, a particular $\mathbb{Z}_2$ liquid~\cite{PunkChowdhurySachdev:Z2} better matches the nearly dispersionless structure %factor measured in Herbertsmithite. Its low-energy spectrum contains flat-band $M_z = 0$ visons and a two-spinon continuum with %both flat and dispersive bands, consistent with the density-of-states peaks and associated $|M_z|=0$ and $|M_z|=1$ bands found %here.

Inelastic spectra of kagome systems are often interpreted as demonstrative of a two-spinon continuum, as incident neutrons cannot excite individual spinons. However, the magnetization-resolved DSSF and $\rho (E)$ [Fig.~\ref{fig:Magnetization_Resolved}a and~\ref{fig:Magnetization_Resolved}b], combined with the robust $\alpha$-scaling of $\omega_1$ and $\omega_2$ to $\alpha = 1$ [Fig.\ref{fig:S_xx_Summary}b], instead suggest that both frequencies are determined by the energies of individual spin-1 excitations.

\tb{Spin gap.}
Additional experimental information that is often invoked to distinguish between possible QSL candidates, e.g., $U (1)$ gapless~\cite{Ran:DSL,Hermele:DSL} or $\mathbb{Z}_2$ gapped~\cite{Sachdev:QMC,YanSimengHuse:DMRG,DepenbrockMcCullochSchollwock:DMRG}, is the absence or presence of a spin gap. We demonstrate in Supplemental Material~\cite{SM} that the $|M_z| = 1$ gaps in $\rho (E)$ found here depend weakly on system size.
Our simulations further demonstrate a gap through the decrease of the elastic response at all momenta with lowering the temperature:
due to the conservation of the quantity
$\sum_\gamma \int_{\bq \in \text{BZ}} \text{d} \bq \int_{-\infty}^\infty \text{d} \omega \, S_{\bq}^{\gamma \gamma} (\omega, \, T)$, such decrease is consistent
with transferring the spectral weight to gapped excitations.

\begin{figure}[t]
    \centering
    \includegraphics[width=0.45\textwidth]{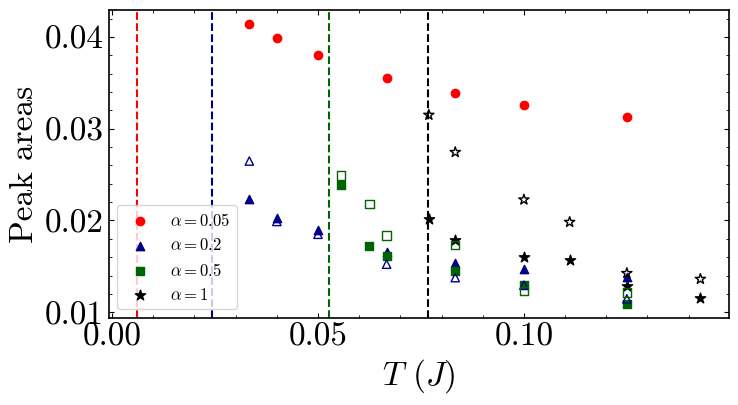}
    \caption{Temperature dependence of the areas $\int_\text{peak} \SF (\omega, \, T) \, \text{d} \omega$ under the low-frequency peaks of $N = 24$ kagome clusters with the indicated values of $\alpha$. For $\alpha = 0.05$, the region of integration spans the broad signal at $0 < \omega \ll 1$. For larger $\alpha$, filled (open) symbols show the area under the peak centered at $\omega_{1(2)}$. Dashed vertical lines show the hidden energy scale $T^* (\alpha)$ corresponding to the same-color set of points.
    \label{fig:Intensity_vs_T}
    }
\end{figure}

\tb{Temperature dependence.} The growth of the low-$\omega$ inelastic intensity upon cooling also reveals the transfer of spectral weight to states at the hidden energy scale. We demonstrate this in Fig.~\ref{fig:Intensity_vs_T} by plotting the temperature dependence of the area of $\SF (\omega, \, T)$ under the low-energy peaks. For $\alpha = 0.05$, the growth of the single low-energy peak captures the growth of the combined intensity of all processes with energy transfers $\Delta E \ll J$. For larger $\alpha$, the shift of intensity from $\omega \approx 0 \rightarrow \omega = \omega_1, \, \omega_2$ shows that most scattering from states at $T^*$ moves to the low-energy inelastic response, while the elastic scattering is due to higher excited states. At all values of $\alpha$, the low-energy peak areas increase monotonically upon approaching the corresponding $T^* (\alpha)$, revealing the transfer of spectral weight.

%%%%%%%% OLD PARAGRAPH %%%%%%%%
% Existing NS data on strongly frustrated materials unanimously show increased low-energy scattering intensity upon cooling below the hidden energy scale $T^*$~\cite{RamirezSyzranov:SRO}. Fig.~\ref{fig:S_xx_Summary} validates this trend. For $\alpha = 0.05$, the growth of the single low-energy peak mimics the growth of the combined intensity of all processes with energy 
% transfers $\Delta E\ll J$.
% For larger $\alpha$, at which three low-energy  peaks can be resolved, the height of the elastic peak grows with $T$, whereas the heights of the peaks at $\omega_1$ and $\omega_2$ decrease with $T$. Thus, most scattering from states at $T^*$ shifts to the low-frequency inelastic response, while elastic scattering receives more contributions from higher excited states.

% As NS experiments often bin intensity over energy-transfer ranges, effectively measuring a frequency-integrated DSSF, Fig.~\ref{fig:Intensity_vs_T} plots the temperature dependence of the area of $\SF(\omega,T)$ under the low-frequency peaks. In agreement with experiment~\cite{Broholm:SRO,Gardner:SRO,Helton:SRO,Stock:SRO,Yang:SRO,LaBarre:SRO,RamirezSyzranov:SRO}, the integrated intensity $\SF(\omega, \, T)$ increases monotonically upon cooling toward the corresponding $T^*(\alpha)$ obtained from $C(T)$. \tr{Should we perhaps move the last phrase to "Comparison with experiments"?}

% the increase of the elastic peak height with $T$ and simultaneous decrease of the peak heights at $\omega_1$ and $\omega_2$ show

%%%%%%%%%%%%%%%%%%%%%%%%%%%%%%%%%%%%%%%%%%%%%%%%%%%%%%%%%%%%%%%%%%%%%

\tb{Comparison with experiments on kagome compounds.}
Considerable effort has examined spin-$1/2$ kagome antiferromagnets using neutron scattering, in both powder~\cite{Helton:SRO,deVries:Herbertsmithite,Helton:Susceptibility} and single-crystal~\cite{Han:Herbertsmithite1,Han:Herbertsmithite2,Zeng:Spinons,Thennakoon:KHAF,Breidenbach:Zn-Barlowite,Jiang:KHAM} samples. For comparison with our numerical results, we focus on single-crystal data.

In Herbertsmithite, a paradigmatic example of a spin-$1/2$ kagome system, $J \approx 17$ meV, and the intensity at $\bq=\Mext$ shows a broad maximum near $\hbar \omega = 5$ meV~\cite{Han:Herbertsmithite1}. This is in good agreement with our result $\omega_1=0.29J$ for $\alpha=1$, previously also reported for a modified DSSF in Ref.~\cite{Prelovsek:DSSF}. However, single-crystal data at $\hbar\omega \gtrsim 0.5J$ are currently unavailable, which 
places the $\omega_2$ peak outside of the experimentally accessed energy range
for Herbertsmithite.

By contrast, neutron spectra for Cs$_8$RbK$_3$Ti$_{12}$F$_{48}$ extend up to 15 meV $\approx 3.7 J$~\cite{Thennakoon:KHAF}. At $\bq=\Mext$, peaks appear near $\hbar \omega = 1$ meV and $2.1$ meV, closely matching our results $\omega_1=0.29J$ and $\omega_2=0.51J$ for $\alpha=1$.

While our simulations also indicate spin gaps associated with $\omega_1$ and $\omega_2$, various experiments point to both gapless~\cite{Han:Herbertsmithite1,Zeng:Spinons,Thennakoon:KHAF} and gapped~\cite{Han:Herbertsmithite2,Fu:Herbertsmithite,Breidenbach:Zn-Barlowite} excitations. 
The non-vanishing sub-gap DoS in some of the experiments
may come from weak disorder~\cite{Han:Herbertsmithite2}. Further-neighbor exchange and Dzyaloshinskii-Moriya interactions~\cite{CanalLacroix:Kagome,Gong:Kagome,Zhu:DSSF,Jiang:KHAM} may also reshape low-energy spectra by inducing gapless order or altering the QSL ground state. Further experimental studies are therefore needed to decisively verify the existence of a spin gap in the pure Heisenberg kagome system.

The neutron scattering data of GFMs unanimously show increased low-energy, short-wavelength scattering intensity upon cooling below the hidden energy scale $T^*$~\cite{Broholm:SRO,Gardner:SRO,Helton:SRO,Stock:SRO,Yang:SRO,LaBarre:SRO,RamirezSyzranov:SRO}. Our results shown in Fig.~\ref{fig:Intensity_vs_T} agree very well with this trend, as the areas under the low-energy peaks in $\SF (\omega, \, T)$ are all seen to increase upon cooling towards $T^*$.

While the numerical values of the energy scales described above are unique to the kagome lattice, the existence of scales observable in NS that are tied to $T^*$ is a generic feature of GFMs. We show this explicitly with calculations for additional GFM models.

\begin{figure}[t]
    \centering
    \includegraphics[width=0.45\textwidth]{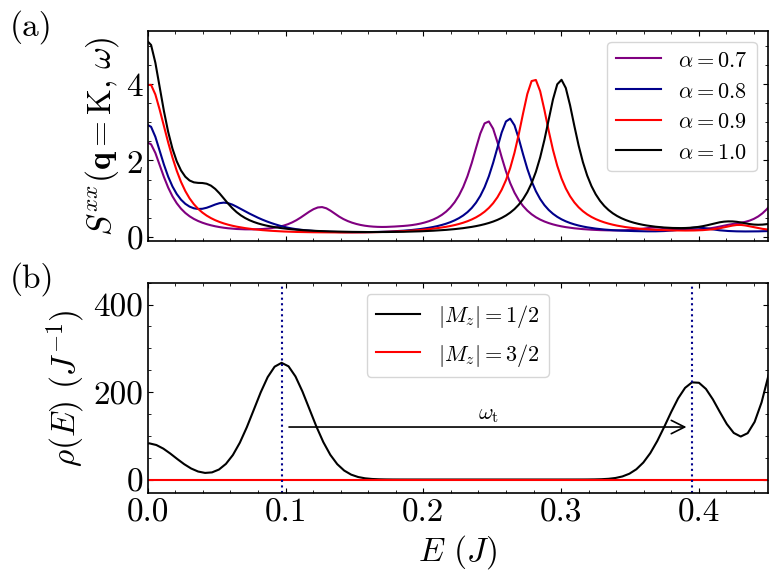}
    \caption{{\bf (a)} $|M_z| = \frac{1}{2} \rightarrow \bigl{\{} \frac{1}{2}, \, \frac{3}{2} \bigr{\}}$ contribution to $S^{xx} (\bq = \text{K}, \, \omega)$ of $N = 27$ triangular clusters at the indicated values of $\alpha$ and $T = J / 15$. {\bf (b)} Low-energy $|M_z| = 1 / 2$ and $|M_z| = 3 / 2$ densities of states for the $\alpha = 1$ cluster.}
    \label{fig:Triangular}
\end{figure}

\tb{Triangular lattice.}
The triangular lattice $J_1 - J_2$ $XXZ$ model is predicted to host a QSL ground state in a finite region of $J_2$ and $\alpha$~\cite{ZhuWhite:THAF,Zhu:THAF,Gallegos:THAF} and underpins numerous candidate QSL materials~\cite{Scheie:THAF1,Scheie:THAF2,Scheie:THAF3,Bag:THAF}. The static structure factor of the QSL is characterized by broad intensity maxima at the K-points of the BZ, reflective of correlations inherited from the 120$^\circ$ ordered state~\cite{PrelovsekKokalj:THAF,Maksimov:THAF,ShermanDupontMoore:THAF,Gallegos:THAF}. We therefore compute the DSSF at $\bq = \text{K}$ of 27-site clusters with $J_2 = 0.1 J_1$ for $\alpha \in [0.7, \, 1]$, along with $\rho (E)$. Results are shown in Fig.~\ref{fig:Triangular}.

As seen in Fig.~\ref{fig:Triangular}a, the DSSF displays a sharp inelastic peak at a frequency $\omega_{\text{t}} \ll J_1$. By computing the heat capacity~\cite{SM}, we verify that $\omega_{\text{t}}$ obeys Eq.~\eqref{eq:triangular_peak}. As shown in Fig.~\ref{fig:Triangular}b, the peak at $\omega_t$ originates from $M_z = \pm \frac{1}{2} \rightarrow \mp \frac{1}{2}$ excited state transitions (the 27-site cluster has ground state $|M_z| = \frac{1}{2}$). As in the case of the kagome lattice (cf. Fig.~\ref{fig:S_xx_Summary}a), an elastic peak also reveals thermally activated transitions between low-lying adjacent magnetization sectors. We note that the growth of the elastic and $\omega_{\text{t}}$ peaks with $\alpha$ at fixed $T$ is due to the growth of the corresponding $\hat{S}_\bq^x$ matrix elements.

\begin{figure}[t]
    \centering
    \includegraphics[width=0.45\textwidth]{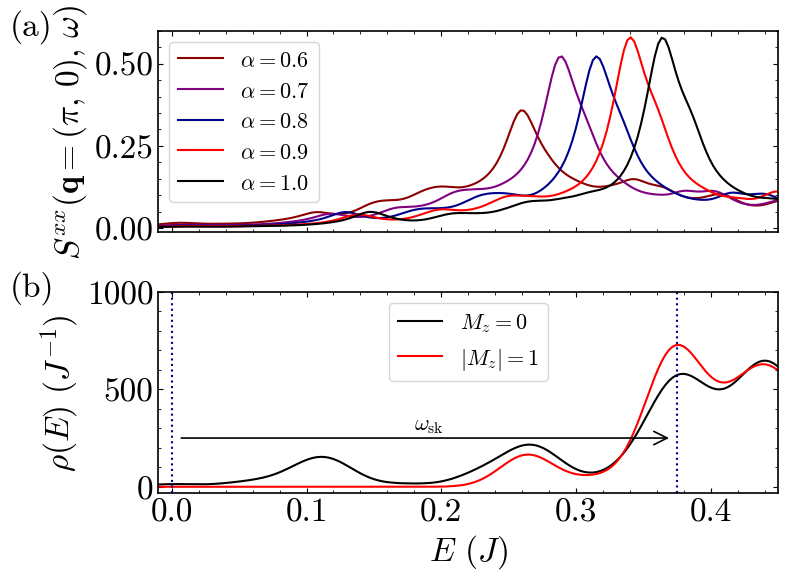}
    \caption{{\bf (a)} $|M_z| = 0 \rightarrow 1$ contribution to $S^{xx} (\bq = (\pi, \, 0), \, \omega)$ of $N = 24$ square-kagome clusters at the indicated values of $\alpha$ and $T = J / 26$. {\bf (b)} Low-energy $|M_z| = 0$ and $|M_z| = 1$ densities of states for the $\alpha = 1$ cluster.}
    \label{fig:Square_Kagome}
\end{figure}

\tb{Square-kagome lattice.}
The square-kagome lattice is a non-Archimedean tiling of corner-sharing triangles with coordination $z = 4$~\cite{SiddharthanGeorges:SK}, and describes numerous QSL candidates of recent interest~\cite{Fujihala:SK,Markina:SK,Gonzalez:SK}. Though there is competing theoretical evidence of a QSL~\cite{LuganJaubertRalko:SK} and valence bond crystal~\cite{Astrakhantsev:SK,Schmoll:SK,JahromiIqbal:SK} ground state, the low-energy manifold and heat capacity are structurally similar to the kagome system~\cite{Richter:SK,RichterDerzhkoSchnack:SK}.
Sample calculations of the DSSF at the extended square BZ corner are shown in Fig.~\ref{fig:Square_Kagome}a. We find a sharp inelastic peak at a frequency $\omega_{\text{sk}} \ll J$, which satisfies Eq.~\eqref{eq:sk_peak} (see Supplemental Material~\cite{SM} for details). As shown in Fig.~\ref{fig:Square_Kagome}b, $\omega_{\text{sk}}$ arises from transitions between the nonmagnetic ground state and $|M_z| = 1$ excited states, similar to $\omega_2$ on the kagome lattice. In contrast to the kagome and triangular cases, there is no elastic response, which we confirm is due to vanishing matrix elements of $\hat{S}_\bq^{x}$ between low-lying $M_z = 0$ and $|M_z| = 1$ states at the simulated momentum. Whether this is true throughout the BZ should be the subject of future study.

%%%%%%%%%%%%%%%%%%%%%%%%%%%%%%%%%%%%%%%%%%%%%%%%%%%%%%%%

\tb{Conclusion.} 
% We have computed finite-temperature dynamical spin correlations of the kagome lattice $XXZ$ antiferromagnet. In the near-Ising limit, the energies at which scattering occurs are determined by the energies of flipping single spins in Ising ground states. For larger exchange anisotropy, the low-energy inelastic scattering is characterized by distinct peaks below the minimum Ising spin-flip energy. By tracking the dependence on anisotropy, we identify the mechanism of the excitations responsible for the low-energy peaks in the structure factor, thereby confirming that they share a common origin with the well-known low-temperature peak in the heat capacity of the kagome antiferromagnet. This suggests a similar route to linking energy scales in different thermodynamic probes for a large variety of GF systems. Magnetization-resolved DSSFs and densities of states indicate that our results for the kagome lattice are consistent with a $\mathbb{Z}_2$ QSL.
{In this Letter, we present a new approach to characterizing the excitations responsible for neutron scattering signatures in GFMs, based on the adiabatic continuity of excitations in the $XXZ$ Heisenberg model. We demonstrate this approach explicitly for the spin-$1 / 2$ kagome, triangular $J_1 - J_2$, and square-kagome lattices. In each case, we find that the finite-temperature dynamical spin structure factor exhibits well-defined inelastic peaks at frequencies well below the exchange coupling. Further, we demonstrate that these frequencies are linearly related to the ``hidden energy scale" $T^*$, which equals the position of a low-temperature peak in the heat capacity, for all values of the exchange anisotropy $\alpha$ for which the corresponding model has a magnetically disordered ground state. The established scaling suggests that the peaks arise from single-magnon excitations, rather than spinon pairs. In addition, on the kagome lattice, we find that the areas under each of the relevant inelastic peaks increase upon cooling $T \rightarrow T^*$, reflective of a universal experimental trend in GFMs.}

% Generally speaking, our results suggest a new, targeted approach to neutron-scattering studies of frustrated materials; particularly, spin-polarized experiments on single crystals, whereby the intrinsic signatures of frustration can be systematically identified.

\tb{Acknowledgments.} Our work has been supported by the DOE Grant No. DE-SC0017862 and the Committee on Research at the University of California Santa Cruz (PP, APR, and SS). Work at Argonne (SR) was supported by the U.S. Department of Energy, Office of Science, Basic Energy Sciences, Materials Science and Engineering. We gratefully acknowledge the computing resources provided on Improv, a high-performance computing cluster operated by the Laboratory Computing Resource Center at Argonne National Laboratory.

% \bibliography{Bibliography}% Produces the bibliography via BibTeX.

\putbib
\end{bibunit}

%%%%%%%%%%%%%%%%%%%%%%%%%%%%%%%%%%%%%%%%%%%%%%%%%%%%%%%%%%%%%%%%%%%%%%%%%%%%%%%%%%%%%%%%%%%%%%%%%%%%%%%%%%%%%%%%
%%%%%%%%%%%%Supplemental Material%%%%%%%%%%%%%%%%%%%%%%%%%%%%%%%%%%%%%%%%%%%%%%%%%%%%%%%%%%%%%%%%%%%%%%%%%%%%%%%

\newpage
\begin{bibunit}
\nocite{apsrev42Control}
\onecolumngrid
\vspace{2cm}
%\twocolumngrid

\cleardoublepage

\setcounter{page}{1}

\renewcommand{\theequation}{S\arabic{equation}}
\renewcommand{\thefigure}{S\arabic{figure}}
\renewcommand{\thetable}{S\arabic{table}}
\renewcommand{\thetable}{S\arabic{table}}
\renewcommand{\bibnumfmt}[1]{[S#1]}
\renewcommand{\citenumfont}[1]{S#1}

\setcounter{equation}{0}
\setcounter{figure}{0}
\setcounter{enumiv}{0}

\begin{center}
	\textbf{\large Supplemental Material for \\
		``Spin correlations, low-energy scales, and anisotropy scaling in 2D frustrated magnets''
	}
	%\\
	%Authors
\end{center}

\section{Differential cross-sections for inelastic neutron scattering}

We consider a spin-polarized neutron beam with polarization axis $\hat{z}$, incident upon a single-crystal sample whose kagome planes are oriented perpendicular to $\hat{z}$. We denote the momentum and spin states of the neutrons as $\{ | \bk, \, \sigma \rangle \}$ and the energy eigenstates of the sample as $\{| p \rangle \}$. The thermal average differential cross-section for scattering from $| \bk_i, \, \sigma \rangle \rightarrow | \bk_f, \, \sigma' \rangle$ is given by~\cite{Squires:TNS}
\begin{equation}
\biggl{\langle} \biggl{(} \DDCS \biggr{)}_{(\bk_i, \sigma) \rightarrow (\bk_f, \sigma')} \biggr{\rangle}_T \propto \frac{k_f}{k_i} \sum_{p, p'} \frac{\mathrm{e}^{-\epsilon_p/ T}}{Z} | \langle \sigma' p' | \ns \cdot \Sperp | \sigma p \rangle |^2 \delta (\omega - [\epsilon_{p'} - \epsilon_p]),
\label{eq:Average_CS}
\end{equation}
where $\bq \equiv \bk_f - \bk_i$, $\Sq$ is the lattice Fourier transform of the sample spin operator, and $\Sperp \equiv \mathbf{q} \times \hat{\mathbf{S}}_{\mathbf{q}} \times \mathbf{q}$. Summing over the initial and final neutron spin states, we obtain the non-spin-flip (NSF) and spin-flip (SF) cross-sections:

\begin{subequations}
    \begin{align}
    \biggl{\langle} \biggl{(} \DDCS \biggr{)}_{(\bk_i \rightarrow \bk_f), \, \text{NSF}} \biggr{\rangle}_T \propto \frac{k_f q^4}{k_i} S_{\bq}^{zz} (\omega, \, T),
    \label{eq:CSNSF}
    \\
    \biggl{\langle} \biggl{(} \DDCS \biggr{)}_{(\bk_i \rightarrow \bk_f), \, \text{SF}} \biggr{\rangle}_T \propto \frac{k_f q^4}{k_i} S_{\bq}^{xx} (\omega, \, T).
    \label{eq:CSSF}
    \end{align}
\end{subequations}

% \begin{figure}[h]
%     \centering
%     \includegraphics[width=0.35\textwidth]{Brillouin_Zones.png}
%     \caption{Elementary (dashed hexagon) and extended (solid hexagon) Brillouin zones of the kagome lattice. M and % $\Mext$ denote the respective M-points. Also shown are the reciprocal lattice vectors $\mathbf{b}_1$ and % $\mathbf{b}_2$ corresponding to the triangular Bravais lattice underlying the kagome network.}
%     \label{fig:Brillouin_Zones}
% \end{figure}

\section{Details of dynamical spin structure factor calculations}
To compute the DSSF~\eqref{eq:DSSF} of spin clusters at finite temperature, we employ a variant of the LTLM~\cite{Aichhorn:LTLM}, described in Refs.~\cite{Kokalj:Thesis,Prelovsek:FTLM}. There are two basic ingredients in the technique. First, the thermal average $\langle \dots \rangle_T$ appearing in Eq.~\eqref{eq:DSSF} is replaced by an average over random initial states:
\begin{equation}
S_\bq^{\gamma \delta} (\omega, \, T) \approx \frac{1}{2 \pi} \int_{-\infty}^{\infty} \mathrm{e}^{i \omega t} \, \frac{\sum_{r = 1}^R \langle r | \mathrm{e}^{-\hat{H} / 2 T} \hat{S}_\bq^{\gamma \dagger} (t) \hat{S}_\bq^\delta (0) \mathrm{e}^{-\hat{H} / 2 T} | r \rangle}{\sum_{r = 1}^R \langle r | \mathrm{e}^{-\hat{H} / T} | r \rangle} \, \mathrm{d} t
\label{eq:DSSF_LTLM}
\end{equation}
where the number $R$ of random states is much smaller than $N_{\mathrm{st}}$, the dimension of the full physical Hilbert space. Second, the expectation values with respect to the random states appearing in Eq.~\eqref{eq:DSSF_LTLM} are evaluated by constructing Krylov subspaces of the Hamiltonian $\hat{H}$, whose dimensions are also significantly smaller than $N_{\mathrm{st}}$. The computational effort of this procedure can be reduced, and the accuracy improved, by using symmetries to block-diagonalize the Hamiltonian. For the $XXZ$ Hamiltonian~\eqref{eq:HXXZ}, we employ translation symmetry and the conservation of $\hat{M}_z$.

Effectively, there are therefore two tuning knobs that control the accuracy of the LTLM: the number $R$ of random states over which the average in Eq.~\eqref{eq:DSSF_LTLM} is taken, and the dimension $M$ of the Krylov subspaces used to calculate the random state expectation values. For $N = 24$ clusters, we use $R \geq 20$ and up to $M = 500$ for symmetry sectors that are too large to effectively fully diagonalize. For $N = 27$ and $N = 30$ clusters (see below), we use $R = 6$ and $M = 300$ ($200$) for $27$ ($30$) site clusters. To obtain the curves shown in Figs.~\ref{fig:S_xx_Summary}a,~\ref{fig:Magnetization_Resolved}a,~\ref{fig:Triangular}a, and~\ref{fig:Square_Kagome}, the discrete spectra obtained from the Krylov subspace diagonalization are convolved with a Lorentzian frequency kernel whose width is chosen according to the typical spacing between poles contributing appreciable weight to the Lehmann representation of Eq.~\eqref{eq:DSSF_LTLM} in the frequency interval of interest.

\begin{figure*}[t]
    \centering
    \includegraphics[width = 0.9\textwidth]{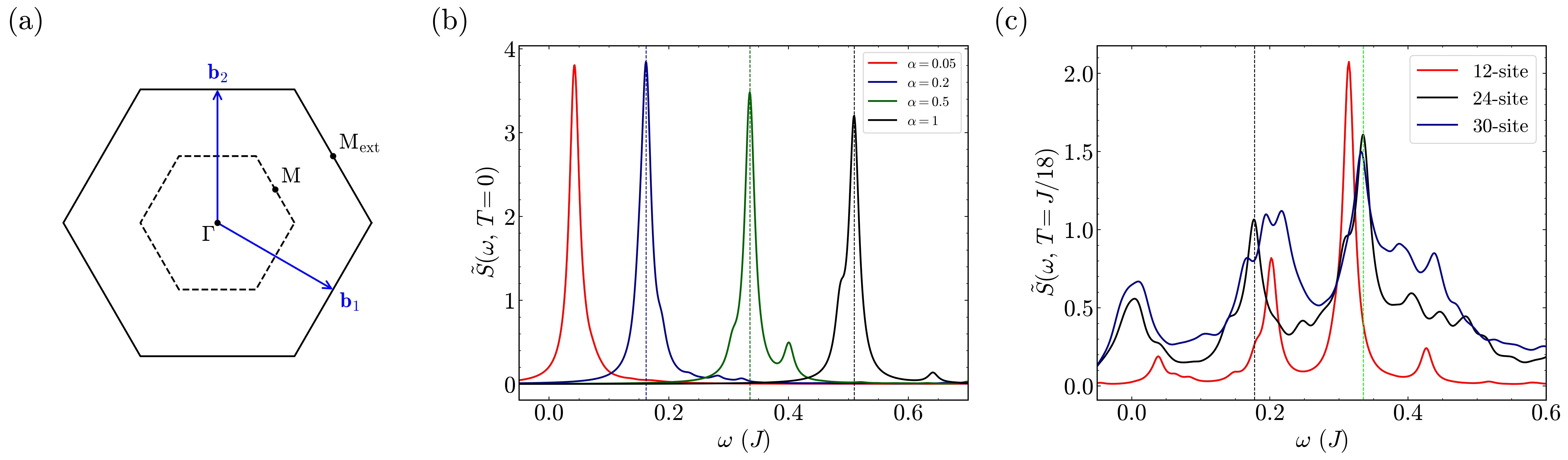}
    \caption{{\bf (a)} Elementary (dashed hexagon) and extended (solid hexagon) Brillouin zones of the kagome lattice. M and $\Mext$ denote the respective M-points. Also shown are the reciprocal lattice vectors $\mathbf{b}_1$ and $\mathbf{b}_2$ corresponding to the triangular Bravais lattice underlying the kagome network. {\bf (b)} Ground-state DSSF for the indicated values of $\alpha$. Dashed vertical lines mark the value of $\omega_2$ in the finite-temperature DSSF at the corresponding $\alpha$. {\bf (c)} Comparison of the $|M_z| = 0 \rightarrow 1$ contribution to $\SF (\omega, \, T = J / 18)$ for different sizes of kagome cluster with $\alpha = 0.5$. Results for $N = 12$ are computed by full ED. Results for $N = 24$ and $N = 30$ are computed by the LTLM. Dashed black (lime) line indicates the position of $\omega_{1 (2)}$ of the 24-site cluster.}
    \label{fig:sm_figure_1}
\end{figure*}

In the $T \rightarrow 0$ limit, the LTLM converges to the ground-state Lanczos method. Results for the $T = 0$ structure factor are shown in Fig.~\ref{fig:sm_figure_1}b.

\section{Finite-size scaling analysis and peak dispersions for the kagome lattice}
\begin{figure}[h]
    \centering
    \includegraphics[width=0.9\textwidth]{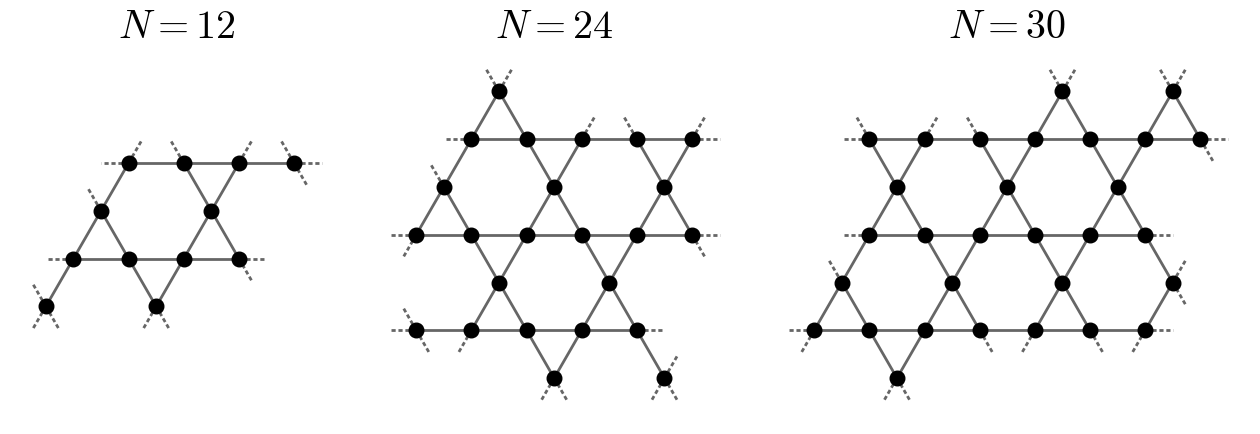}
    \caption{The kagome lattice clusters considered in this Letter.}
    \label{fig:kagome_clusters}
\end{figure}

To demonstrate that the main results of our simulations are robust against finite-size effects, we compute the $|M_z| = 0 \rightarrow 1$ contribution to $\SF (\omega, \, T)$ of $N = 30$-site clusters for a limited number of $(\alpha, \, T)$ points. Representative results are shown in Fig.~\ref{fig:sm_figure_1}c, alongside full exact diagonalization (ED) results for an $N = 12$ cluster. For convenience, the clusters considered in this work are drawn in Fig.~\ref{fig:kagome_clusters}.

% \begin{figure}[h]
%     \centering
%     \includegraphics[width=0.5\textwidth]{Finite_Size_Analysis.png}
%     \caption{Comparison of the $|M_z| = 0 \rightarrow 1$ contribution to $\SF (\omega, \, T = J / 18)$ for % different sizes of kagome cluster with $\alpha = 0.5$. Results for $N = 12$ are computed by full ED. Results % for $N = 24$ and $N = 30$ are computed by the LTLM. Dashed black (lime) line indicates the position of % $\omega_{1 (2)}$ for the 24-site cluster.}
 %    \label{fig:Finite_Size}
% \end{figure}

Remarkably, the two-peak behavior is evident even for $N = 12$, albeit with slightly shifted (within $0.03 J$) positions of the peaks relative to the $N = 24$ case. At $N = 30$, the position of $\omega_2$ is virtually unchanged from $N = 24$, falling within $0.002 J$. The behavior near $\omega_1$ is somewhat more ambiguous, with 3 closely-spaced peaks appearing within $0.05 J$ of the value of $\omega_1$ obtained for $N = 24$. The highest of these peaks falls within $0.02 J$. Given the much smaller sampling that we can afford for $N = 30$ ($R = 6$ vs. $R \geq 20$), it is possible that the observed structures near $\omega_1$ for $N = 30$ are due to sampling uncertainty, though such uncertainties are reduced for larger clusters. Alternatively, it is possible that $M = 200$-state Krylov subspaces are insufficient to resolve a single smooth peak at $\omega_1$. Nonetheless, the robustness of the qualitative structure of the DSSF and the positions of the low-energy peaks across the range of simulated sizes are strong indications that the results presented in this Letter reflect the true physics of the kagome antiferromagnet.

Additional quantities of interest in the finite-size scaling of our results are the value of $T^*$, the energy $\Delta_0$ of the lowest peak in the $|M_z| = 0$ density of states and the energies $\Delta_1^{(1)}$ and $\Delta_1^{(2)}$ of the two lowest peaks in the $|M_z| = 1$ density of states, and the values of the coefficients $A_{1(2)}$ in the relation $\omega_{1(2)} = A_{1(2)} T^*$. The values of these quantites, along with the peak frequencies $\omega_1$ and $\omega_2$, are summarized in Table~\ref{tab:finite_size_summary}.

\begin{figure}
\centering
    \includegraphics[width = 0.9\textwidth]{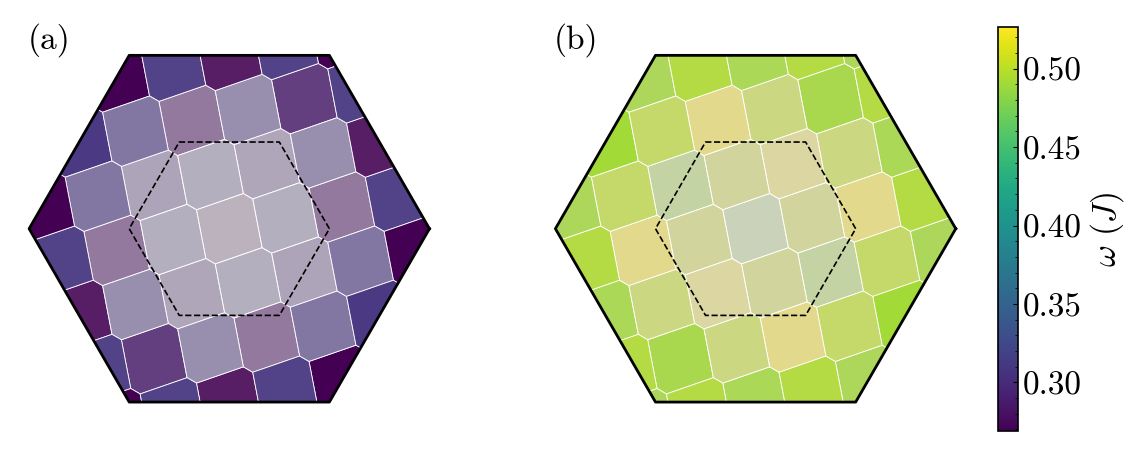}
    \caption{Dispersion of {\bf (a)} $\omega_1 \approx 0.29 J$ and {\bf (b)} $\omega_2 \approx 0.51 J$ throughout the extended BZ for $\alpha = 1$, $T = J / 13$. Panel opacity indicates the magnitude of $S_{\bq}^{xx} (\omega, \, T)$ at the corresponding momentum $\bq$. Dashed hexagons indicate the boundary of the elementary BZ.}
    \label{fig:sm_figure_2}
\end{figure}

\begin{table}[h]
    \centering
    \begin{ruledtabular}
    \begin{tabular}{ccccccccc}
         Cluster size & $\omega_1$ & $\omega_2$ & $T^*$ & $A_1$ & $A_2$ & $\Delta_0$ & $\Delta_1^{(1)}$ & $\Delta_1^{(2)}$ \\
         \hline
         12 & $0.33$ & $0.46$ & $0.11$ & $2.98 \pm 0.01$ & $4.35 \pm 0.05$ & $0.139$ & $0.382$ & $0.584$ \\
         24 & $0.29$ & $0.51$ & $0.077$ & $3.40 \pm 0.09$ & $6.236 \pm 0.012$ & $0.109$ & $0.375$ & $0.512$ \\
         30 & $0.31$ & $0.50$ & $0.081$ & $3.82 \pm 0.11$ & $6.42 \pm 0.16$ & $0.086$ & $0.375$ & $0.540$ \\
    \end{tabular}
    \end{ruledtabular}
    \caption{Cluster size-dependence of the key quantities discussed in this work for the kagome lattice. All energy scales are computed for the $\alpha = 1$ cluster, and given in units of $J$. The coefficients $A_1$ and $A_2$ are dimensionless, and obtained by fitting simulated data for multiple values of $\alpha$ to $\omega_{1(2)} (\alpha) = A_{1(2)} T^* (\alpha)$.}
    \label{tab:finite_size_summary}
\end{table}
Another key result of our simulations for the kagome lattice is the robustness of the peak positions $\omega_1$ and $\omega_2$ throughout the extended BZ for a fixed value of $\alpha$. This is shown in Fig.~\ref{fig:sm_figure_2}.

\section{Heat capacities and structure factor peak fitting of triangular $J_1 - J_2$ and square-kagome clusters}

\begin{figure}[h]
    \centering
    \includegraphics[width = 0.9\textwidth]{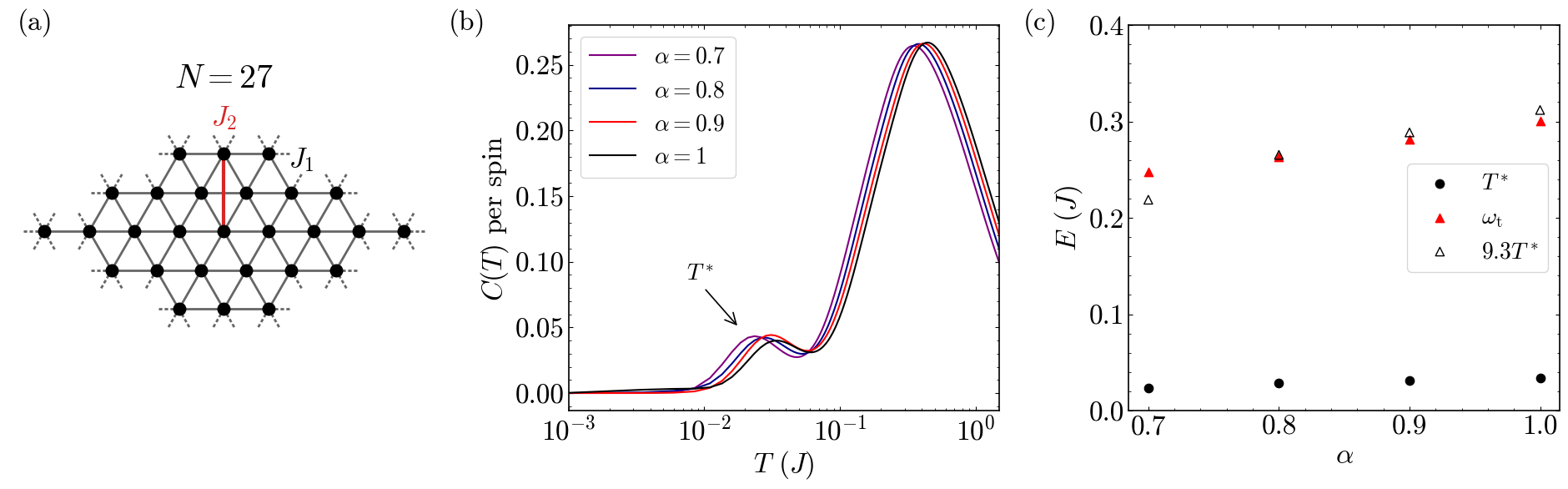}
    \caption{{\bf (a)} The $N = 27$ triangular cluster used in this work. $J_1$ and $J_2$ labels show nearest and next-nearest neighbor bonds, respectively. {\bf (b)} Heat capacity of the $N = 27$ cluster at the indicated values of $\alpha$. {\bf (c)} Dependencies of the energy scales $T^*$ and $\omega_{\text{t}}$ on $\alpha$ (cf. panel {\bf b}).}
    \label{fig:Triangular_SM}
\end{figure}

\begin{figure}[h]
    \centering
    \includegraphics[width = 0.9\textwidth]{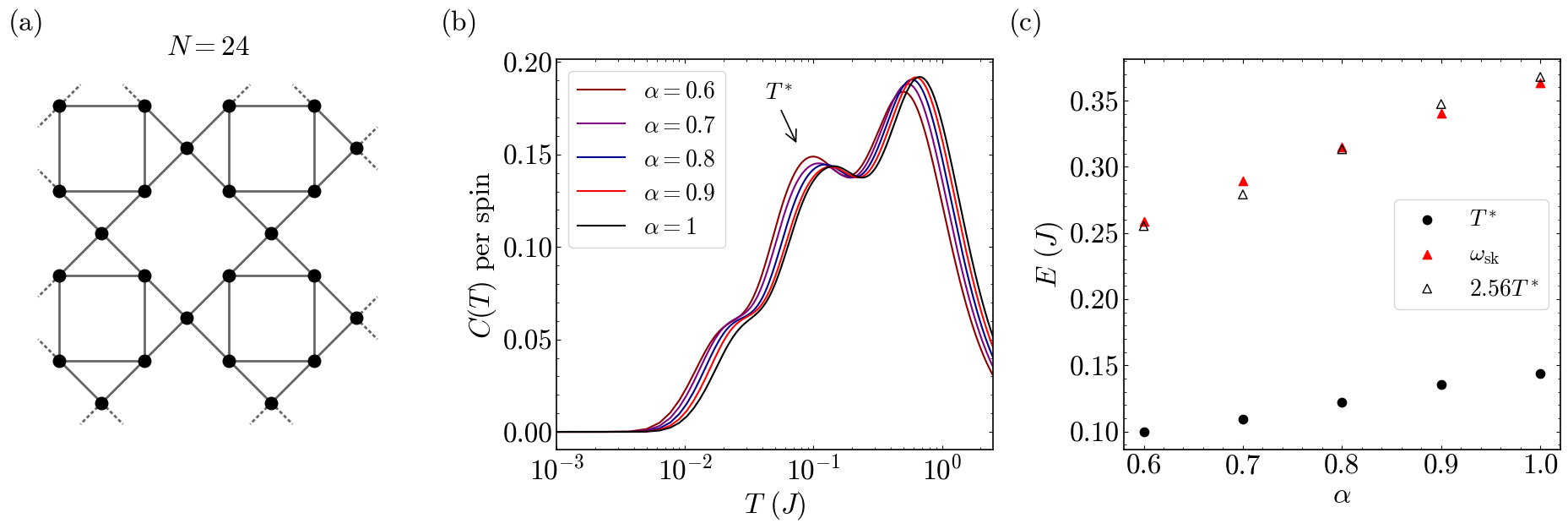}
    \caption{{\bf (a)} The $N = 24$ square-kagome cluster used in this work. {\bf (b)} Heat capacity of the $N = 24$ cluster at the indicated values of $\alpha$. {\bf (c)} Dependencies of the energy scales $T^*$ and $\omega_{\text{sk}}$ on $\alpha$ (cf. panel {\bf b}).}
    \label{fig:Square_Kagome_SM}
\end{figure}

The relations Eqs.~\eqref{eq:triangular_peak} and~\eqref{eq:sk_peak} for the triangular and SK systems, respectively, are obtained analogously to the relations~\eqref{eq:kagome_peaks} for the kagome lattice: first, the heat capacity is computed by the finite-temperature Lanczos method~\cite{Prelovsek:FTLM} to extract the value of $T^*$. Then, the coefficient relating the corresponding structure factor peak frequency to $T^*$ is identified by a simple linear fit. The triangular and SK clusters considered in this work, along with their heat capacities and structure factor frequency fits, are shown in Figs.~\ref{fig:Triangular_SM} and~\ref{fig:Square_Kagome_SM}, respectively.

%\subsection{Checkerboard lattice}

%\begin{figure}[h]
%    \centering
%    \includegraphics[width = 0.9\textwidth]{sm_figure_3.png}
%    \caption{{\bf (a)} Magnetization-resolved transverse DSSF and {\bf (b)} %density of states of an $N = 24$ checkerboard cluster with $J_2 = \alpha %= 1$. Panel {\bf (a)} was computed at $T = J / 15$. Dotted blue lines in %panel {\bf (b)} indicate the energies of states contributing to the two %inelastic peaks at $\omega_{1 \text{c}}$ and $\omega_{2 \text{c}}$.}
%\end{figure}

%\subsection{Triangular lattice}

%\begin{figure}[h]
%    \centering
%    \includegraphics[width = 0.9\textwidth]{sm_figure_4.png}
%    \caption{{\bf (a)} Magnetization-resolved transverse DSSF and {\bf (b)} %density of states of an $N = 27$ triangular cluster with $J_2 = 0.1$, %$\alpha = 1$. Panel {\bf (a)} was computed at $T = J / 15$. Dotted blue %lines in panel {\bf (b)} indicate the energies of states contributing to %the three inelastic peaks at $\omega_{1 \text{t}}$, $\omega_{2 %\text{t}}$, and $\omega_{3 \text{t}}$.}
%\end{figure}

% \begin{figure}[h]
%    \centering
%    \includegraphics[width = 0.6\textwidth]{Heat_Capacity_Summary.png}
%    \caption{Heat capacity $C (T)$ of 24-site $XXZ$ kagome clusters, computed by the FTLM.
%    \tr{Since this figure is now in the main text, perhaps we don't need here?}
%    }
%    \label{fig:Heat_Capacity_SM}
%\end{figure}

\putbib
\end{bibunit}

\end{document}